\documentclass[12pt]{article}
\usepackage{authblk}
\usepackage[T2A]{fontenc}
\usepackage[main=english,russian]{babel}
\usepackage{cite}
\usepackage{mcite}
\usepackage{amssymb}
\usepackage{geometry}
\usepackage{color,graphicx}
\usepackage{amsmath}
\usepackage{mathtools}
\usepackage{tikz}
\usetikzlibrary{automata, topaths}
\usepackage{wrapfig}
\usepackage{graphicx}
\usepackage{caption}
\usepackage{subcaption}
\usepackage[export]{adjustbox}

\usepackage[colorinlistoftodos]{todonotes} 
\usepackage[colorlinks=true, allcolors=blue]{hyperref}

\usepackage[capitalise]{cleveref}

\definecolor{blue2}{RGB}{76,0,152}

\title{Quantum-field multiloop calculations in critical dynamics}

\author[1]{Ella Ivanova \thanks{ella.v.ivanova@gmail.com}}
\author[2]{Georgii Kalagov \thanks{kalagov.g@gmail.com}}
\author[3]{Marina Komarova \thanks{frommarina@mail.ru}}
\author[2,3]{Mikhail Nalimov \thanks{ m.nalimov@spbu.ru.}}
\affil[1]{ Otto H. York Department of Chemical and Materials Engineering, New Jersey Institute of Technology, University Heights, Newark, NJ 07102, USA}
\affil[2]{Bogoliubov Laboratory of Theoretical Physics, Joint Institute for Nuclear Research, Joliot-Curie 6, Dubna, Moscow region, 141980, Russia}
\affil[3]{Department of Physics,  St. Petersburg State University,  Universitetskaya nab.7/9, St. Petersburg, 199034, Russia}

\begin{document}

\maketitle

\begin{abstract}
{The quantum-field renormalization group method is one of the most efficient and powerful tools for studying critical and scaling phenomena in interacting many-particle systems. The multiloop Feynman diagrams underpin the specific implementation of the renormalization group program. In recent years, multiloop computation has had a significant breakthrough in both static and dynamic models of critical behavior. In the paper, we focus on the state-of-the-art computational techniques for critical dynamic diagrams and the results obtained with their help.  
The generic nature  of the evaluated physical observables in a wide class of field models is manifested in the asymptotic character of perturbation expansions. Thus, the Borel resummation of series is required to process multiloop results. Such a procedure also enables one to take high-order contributions into consideration properly.
The paper outlines the resummation framework in dynamic models and the circumstances in which it can be useful.
An important resummation criterion is the properties of the higher-order asymptotics of the perturbation theory. In static theories, these properties are determined by the method of instanton analysis. A similar approach is applicable in critical dynamics models. We describe the calculation of these asymptotics in dynamical models and present the results of the corresponding resummation.}
\end{abstract}

%%%%%%%%%%%%%%%%%%%%%%%%%%%%%%%%%%%%%%%%%%
\section{Introduction\label{sec:intro}}

Continuous  phase transitions controlled by an order parameter play an essential role in the description of condensed systems. This applies equally to the well-known phase transitions, which have been included in textbooks for hundreds of years, and to the exotic ones, obtained relatively recently in powerful fields or at superhigh pressures. A wide variety of critical points are known: these points represent a set of thermodynamic parameters near which the behavior of the system is non-analytical. The anomalies demonstrate ``universality'', expressed, for example, in the existence of universal critical indices. The latter is determined by rather general properties of the system under consideration, for example, by the tensor structure of the order parameter and space dimension.

Despite a rich history, the theory of critical behavior began to develop only in the 20th century, when L.D.~Landau proposed the simplest description of continuous phase transitions based on the analysis of symmetry and dimension \cite{Landau1937,Landau21937}. However, Landau's theory was based on an incorrect account of fluctuations in the considered systems and therefore gave incorrect predictions. Significant progress in the theory of critical phenomena was associated with the emergence of the so-called Wilson renormalization group (RG) \cite{Wilson1974,Patashinskij1979,Ma2000} and the $\varepsilon$-expansion. Here, for the first time, it was possible to take into account the renormalization of critical exponents due to fluctuation effects.
Similar techniques have yielded  both the static properties of the critical behavior and the dynamic  properties of equilibrium fluctuations, which describe the growth of the relaxation times  in the vicinity of the critical point. Unfortunately, the consistent application of the Wilsonian RG turned out to be associated with considerable computational difficulties. Therefore, in this computational formalism, it was only possible to obtain significant results within the scope of the first-order perturbation theory. Of course, such  accuracy did not correspond to experimental expectations and did not allow to make practically important predictions, which the theory of condensed matter so badly needed.
Fortunately, it turned out that the equivalent and more advanced technique of quantum field RG \cite{Zinn1996,Klinert2001,Vasilev2000} solves the problem much better. On its basis, the development of theoretical studies of critical behavior continued.

In the theory of critical behavior, there are quite a few different renormalization group approaches. Among them, in addition to the Wilsonian RG mentioned above, we will name the RG in the real space dimension \cite{PhysRevB.17.1365,Orlov2000,Adzhemyan2016} as well as the functional RG \cite{Dupuis2021}. Our article is mainly devoted to the standard quantum field RG approach. It allows the calculation of multiloop diagrams most efficiently with proper use of the arbitrariness of renormalization scheme. The result for critical exponents is constructed as a series in $\varepsilon$ parameter, which is the deviation of the space dimension from the logarithmic dimension,  for which the Lagrangian coupling constant has specific scaling properties. In presenting the principles of multiloop calculations, we assume that the reader is familiar with quantum field renormalization group methods in statistical physics as well as with the theory of ultraviolet renormalization.  For our part, we recommend here the classic monographs \cite{Zinn1996, Vasilev2000, Klinert2001}. 
In the process of time-consuming multiloop calculations, it turned out that progress is influenced not only by technical methods but also by the regularization and renormalization schemes used. For dimensional regularization, the minimal subtraction scheme in the massless theory is the most efficient (see, for example, \cite{Vasilev2000} and references there). With this approach, the best results of multiloop calculations in critical dynamics are achieved.
 
In this article, we will refer to the results obtained in the study of ``critical dynamics''. Here we have in mind the dynamics of fluctuations near the equilibrium state in the vicinity of the critical point. It should be noted that the presented methods for calculating diagrams and for resummation of perturbation series are also successfully applied to models studied by the renormalization group in the real space dimension and to microscopic quantum models. Furthermore, they are also successfully used in various stochastic models that do not tend to the Gibbs equilibrium in the limit of long relaxation time. However, these models still need to sufficiently implement  multiloop calculations.

The article is organized as follows. Section \ref{sec:calc} outlines the main diagram calculation techniques developed for static field theories. The basic models of critical dynamics are described in section \ref{sec:dyn_models}. Section \ref{sec:calc_dyn} presents the currently developed methods for calculating dynamics diagrams and a simple illustration of some calculations. The fluctuation-dissipation theorem leads to the connections between the static and corresponding dynamics models. This leads to significant simplifications in the multiloop dynamics calculations. This is a basis of diagram reduction for the critical dynamics models described in section~\ref{sec:red}, where the result of five-loop calculations for the  $O(n)$ symmetric model $A$ is presented. Section~\ref{sec:borel} is devoted to methods of resuming divergent series in critical dynamics models. The resummed values of dynamic critical exponents $z$ for the model $A$ are presented.

\subsection{\label{sec:calc}Basic diagram calculation methods}

The standard  $\phi^4$ model, illustrating multiloop calculation methods, is considered as the theory of the scalar field $\phi(x)$ over the Euclidean space of dimension $d=4-\varepsilon$. The system is described by the action
\begin{align} 
\label{eq:action}
    S =\int \mathrm{d}^{d} x \left( \frac{1}{2} (\nabla\phi)^2 +\frac {\tau_0}2\phi^2+\frac {g_0}{4!}\phi^4\right).
\end{align}
As is customary in the literature, in similar expressions below, we will omit integration over spatial coordinates and time. Although the investigated functional has the meaning of a Hamiltonian, in the quantum field approach, it is denoted by $S$, inheriting the notation from the relativistic QFT in the Minkowski space. The model under consideration is multiplicatively renormalizable. The renormalization is performed by stretching the fields $\phi\to Z_{\phi}\phi$ and the parameters $g_0=g\mu^{\varepsilon}Z_g$, $\tau_0=\tau Z_{\tau }$, where $\mu$ is the renormalization mass. In the minimal subtraction scheme, the renormalization constants have purely pole contributions generated by divergent diagrams 
\begin{equation}
Z_i=1+\frac {c^{(1)}_{i}(g)}{\varepsilon}+\frac {c^{(2)}_{i}(g)}{\varepsilon^2 }+\dots, \quad i \in \{\phi, g, \tau \}.
\end{equation}
The method of studying critical behavior is based on the RG equation (see, for example, \cite{Vasilev2000} and references therein)
\begin{equation*}
 [s \partial _{s}-\beta \partial _g+ \Delta_{\tau}\tau\partial_{\tau}+\Delta _n]G_n^R=0,
\end{equation*}
where $s=p/\mu$ is the scale parameter and $G_n^R$ is the $n$-point correlation function. The so-called renormalization group functions $\beta$ and $\Delta _i=d_i+\gamma _i$ (here $d_i$ are the canonical dimensions of $i$, $\gamma_i$ correspond to the anomalous dimensions) are expressed in a known way in terms of the renormalization constants \cite{Vasilev2000}. Small $s$ corresponds to the large-scale-- or infra-ref (IR)--behavior of the system's correlators. The IR region is reached at $p\to 0$, for example, when $p$ is the only argument of the correlation function in the momentum representation.
Thus, the standard problem in this approach is to calculate the contributions of all diagrams to the renormalization constants for a given order of the perturbation expansion. Further, a fairly routine recalculation of them into the $\varepsilon$-expansion coefficients of critical exponents is required. Let us list the main techniques for calculating diagrams that have been developed in statics field theories \cite{Vasilev2000,Smirnov2006}.

 \textit{Alpha-representation.} It consists of the integral representation of a power factor in the denominator of the diagram in the form 
\begin{equation}
\label{eq:alpha}
\frac1{a^{\alpha}}=\frac 1{\Gamma(\alpha)}\int\limits_0^{\infty}\mathrm{d}t \, t^{\alpha-1}e^{-a t }\, ,
\end{equation}
and the subsequent replacement of the orders of integration in the diagram. As a result, the $d$-dimensional integrals over the momenta are calculated first, and the parametric integrals are processed at the end.

 \textit{Feynman representation.}
\begin{equation}
\label{eq:feyn} \frac1{a_1^{\alpha_1} \dots a_n^{\alpha_n}}=\frac {\Gamma(\alpha_1 + \dots + \alpha_n)}{\Gamma(\alpha_1)\dots\Gamma(\alpha_n)}\int\limits_0^{1}\dots \int\limits_0^{1}\prod\limits_{j=1}^{n} \mathrm{d} z_j\frac{z_j^{\alpha_j-1}\delta(\sum_{i=1}^{n} z_i-1)}{(\sum_{i=1}^{n} z_i \, a_i)^{\alpha_1+\dots+\alpha_n}}.
\end{equation}
Let's consider an example of applying the mentioned techniques to calculate the pole singularity of the following diagram in the dimensional regularization $d = 4 - \varepsilon$
\begin{equation}
 \includegraphics[scale=0.5,valign=c]{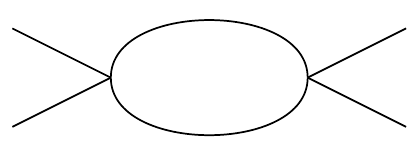}=\int\frac {\mathrm{d}^d k}{(2\pi)^d}\frac{1}{[k^2+\tau][(p-k)^2+\tau]}=\int\frac {\mathrm{d}^d k} {(2\pi)^d}\int\limits_0^{1} \mathrm{d}z\frac 1{[(k^2+\tau)z+(p-k)^2(1-z)]^2}.
\end{equation}
The calculation of the last expression is simplified if we recall the independence of the calculated pole contribution in $\varepsilon$ from the external momentum $p$. Setting $p=0$ and using the $\alpha$-representation, we get
\begin{align*}
    \includegraphics[scale=0.5,valign=c]{fig/ryba.pdf}
 =\int\frac {\mathrm{d}^d k} {(2\pi)^d}\int\limits_0^{1} \mathrm{d}z\int\limits_0^{\infty} \mathrm{d}t \, t e^{-t(k^2+\tau)}=\frac{\pi^{d/2}}{(2\pi)^d}&\int\limits_0^{\infty} \mathrm{d}t \, t^{1-d/2} e^{-t \tau} \\& =\frac{\pi^{d/2}}{(2\pi)^d}\tau^{d/2-2}\Gamma(2-d/2).
\end{align*}
The pole contribution in $\varepsilon$ in this expression does not depend on $\tau$, which is the basis for using the ``massless'' calculation scheme with $\tau =0$.
In more complex diagrams, after calculating the $d$-dimensional integrals over the arguments of the propagators, a non-trivial problem is the evaluation of the integrals over the remaining set of parameters. Let's designate this set of parameters $z_i$ and $t_i$ and list the main techniques employed to compute them.

 \textit{Sector decomposition method.} Within this machinery \cite{Heinrich2012} the parametric multiple integral is divided into sectors $z_1\ge\{z_2,\dots ,z_n\}$, $z_2\ge\{z_1, z_3,\dots ,z_n\}$, etc., and then the appropriate scaling of variables $z_i$ is performed in each sector. As a result, all integrals again take the form $\int\limits_0^{1} \dots \int\limits_0^{1}\prod \mathrm{d}z_i$. The integral over the large parameter $z$ is easy to calculate. It produces poles in $\varepsilon$. The sector decomposition procedure can be repeated for diagrams with divergent subgraphs to obtain a convergent integral.
 
 \textit{Kompaniets-Panzer method.} The method \cite{Kompaniets2017a} 
 is based on the fact that integrals over parameters $t_i$ of alpha-representations in the diagrams without divergent subgraphs or in the subgraphs can be calculated with a certain order of integration using analytical formulas for polylogarithms.

 \textit{Alternating transitions to the coordinate and momentum representations.}
A significant simplification in analytical calculations was given by the method presented in \cite{Vasilev2000}, using formulas for the Fourier transform $\mathcal{F}$ of arbitrary power functions in an arbitrary space dimension
\begin{equation}
\frac{1}{(x_1-x_2)^{2 \, \alpha}} \quad \xRightarrow[]{\mathcal{F}} \quad  \frac{c(\alpha)}{k^{2\,(d/2-\alpha)}}    
\end{equation}
 and formulas for the inverse Fourier transform $\mathcal{F}^{-1}$
\begin{equation}\frac 1{k^{2\alpha}} \quad \xRightarrow[]{\mathcal{F}^{-1}} \quad \frac{ \bar c(\alpha)}{(x_1-x_2)^{2(d/2-\alpha)}}.
\end{equation}
Here we introduce magnitudes $$c(\alpha)=\pi^{d/2}4^{d/2-\alpha}\Gamma(d/2-\alpha)/\Gamma(\alpha),
\qquad
\bar c(\alpha)=c(\alpha)/(2\pi)^d.$$
For example, calculating a two-loop diagram using this technique looks like this
\begin{equation}
\includegraphics[scale=0.2,valign=c]{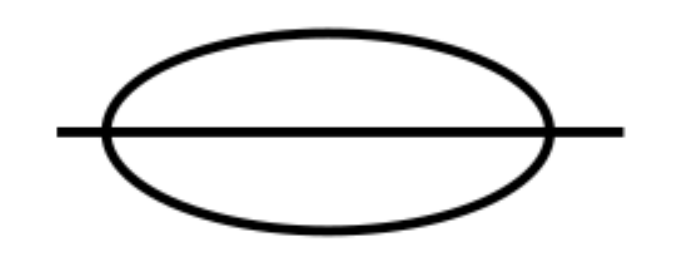}= 
\int\frac{\mathrm{d}^d k \,\mathrm{d}^d q}{(2\pi)^{2d}}\frac 1{k^2 \,q^2\,(p-k-q)^2}  \, \xRightarrow[]{\mathcal{F}^{-1}}\frac{{\bar c}(1)^3}{ (x_1-x_2)^{2(3d/2-3)}} \, \xRightarrow[]{\mathcal{F}} \,\frac{ {\bar c}(1)^3 \, c(2 d/3-1)\,} {p^{2(3-d)}}.
\end{equation}
This expression explicitly contains the required $\varepsilon$ pole contributions. For calculating more complex diagrams, the convolution of lines formula is also useful:
\begin{equation}
\label{eq:convst}
\int \mathrm{d}^d{x} \,\frac{1}{({ x}_1-{ x})^{2\alpha_1}} \, \frac 1{({ x}_2-{ x})^{2\alpha_2}}= \frac {c(\alpha_1)c(\alpha_2)\bar c(d-\alpha_1-\alpha_2)}{({x}_1-{x_2})^{2(\alpha_1+\alpha_2-d)}}.
\end{equation}

Further development of this method was based on the use of the vertex uniqueness method \cite{Kazakov1983} and the $R^*$-operation \cite{Chetyrkin1982,Chetyrkin1983,Chetyrkin1981,erChetyrkin1981}. The described methods made it possible to evaluate the critical exponents in the $\phi^4$ static model up to six-loop diagrams \cite{Kompaniets2017a}, and present the critical indices in the form of the $\varepsilon$-expansion up to the corresponding order. As a rule, the $\varepsilon$-series obtained in the static models turns out to be asymptotic, so the resulting segment of the series required a resummation procedure.

\section{\label{sec:dyn_models}Models of critical dynamics.}

The study of static properties of correlated equilibrium systems in the vicinity of the critical point or close to the continuous phase transitions 
is a major area of statistical field theory. However, time-dependent phenomena, e.g., relaxation dynamics, are more diverse and no less interesting because they capture inter-mode interaction effects, which die out in equilibrium. A key experimental manifestation of 
dynamic effects is critical slowing down in the 
relaxation time, which increases abnormally. The most investigated and known models of 
critical dynamics are listed in \cite{Hohenberg1977} and in more detail discussed in
\cite{Tauber2012,Vasilev2000, Folk2006}.

Critical dynamics is described by a set of physically meaningful fields $\phi_a$, which obey the system of stochastic differential equations. For a wide class of systems, these equations can be written in the general Langevin form 
\begin{align}
\label{eq:main_eq}
\partial _t \phi_a =-(\alpha_{ab}+\beta_{ab})\frac{\delta S}{ \delta {\phi_b}}+\eta_a, 
\end{align}
where  $\alpha$ is a symmetric operator $\alpha_{ a b} = \alpha_{b a} $, $\beta$ is a skew-symmetric one $\beta_{ a b} = -\beta_{b a} $ and satisfy the condition ${\delta \beta_{ab}}/{\delta \phi_a}=0$, where summation over the index $a$ is assumed \cite{Vasilev2000}. It may be verified that this condition is sufficient to ensure relaxation towards an equilibrium state with the Gibbs distribution functional $\exp(-S)$. The additive noise $\eta_a$ is the Gaussian random field and is fully characterized by the first two moments 
\begin{align}
\langle \eta_a(x,t) \rangle = 0, \quad
\langle \eta_a(x,t) \eta_b(x',t') \rangle = 2 \alpha_{a b} \,\delta^d(x-x')  \delta(t-t').
\end{align}
The explicit form of the static action $S$, which encodes the types of underlined symmetry and strengths of the assorted interactions, depends on the phase transitions or critical points we explore.  

Moving on, let us review the paradigmatic models which are at the heart of a tremendous number of research papers on critical dynamical behavior. The simplest one is {\it the model $A$} that encompasses the critical dynamic of a single non-conserved order parameter. The corresponding action is given by
\begin{equation}
\label{eq:S1}
S =\frac{1}{2}(\nabla\phi)^2+\frac {\tau}2\phi^2+\frac{g}{4!}\phi^4, \qquad
\alpha=const >0.
\end{equation}

{\it The  model $B$} is determined by the same static action, \cref{eq:S1}, and describes relaxation of the conserved field, resulting in the differential operator ${\alpha}=-\lambda \nabla^2$ with a positive kinetic coefficient $\lambda>0$.

The rich collective dynamics of a non-conserved field $\phi$ and conserved scalar one $m$ are captured by the effective action of {\it the model $C$}
\begin{equation}
\label{eq:S2}
S =\frac{1}{2}\,(\nabla\phi)^2+\frac{g_{\phi}}{4!}\phi^4+\frac{m^2}{2}+ \frac{1}{2}\,g_{m \phi} \,m \, \phi^2\, , \qquad {
\alpha}= \left(
\begin{array}{cc}
\lambda_{\phi}& 0\\
0& -\lambda_{m} \nabla^2
\end{array}
\right)\, ,
\end{equation}
with a positive set of coefficients $\{\lambda_{\phi},\lambda_{m}\}$ and two independent coupling constants  $\{g_{\phi},g_{m \phi}\}$.

The study of fully conserved dynamic of two fields $\phi$ and $m$ with the equilibrium action, \cref{eq:S2}, gives rise to {\it the model $D$}, where the operator $\alpha$ has the form 
\begin{equation}
\alpha= \left(
\begin{array}{cc}
-\lambda_{\phi} \nabla^2& 0\\
0& -\lambda_{m} \nabla^2
\end{array}
\right).
\end{equation}
Until now, the above definitions of the stochastic models have not incorporated all the possibilities of the entire equation, \cref{eq:main_eq}. One can specify the operator $\beta$ responsible for the inter-mode interaction during the relaxation process. On the one hand, additional couplings $\beta$ lead to a more diverse physical picture. On the other hand,  this sufficiently complicates the theoretical analysis, which is reflected in the absence of multi-loop renormalization group calculations, and the available results that some authors led are currently controversial. 

The static action of {\it the model F} of the complex valued field $\psi$ and the real scalar one $m$  is 
\begin{equation}
S = \nabla\psi \nabla\psi^+ + \frac{g_{\psi}}{3!}(\psi \psi^+)^2+ \frac{m^2}{2}+ g_{m \psi} \, m \psi \psi^+,
\end{equation}
whilst 
\begin{eqnarray}
&& {\alpha}= \left(
\begin{array}{ccc}
0&\lambda _{\psi}& 0\\
\lambda _{\psi}&0&0\\
0&0& -\lambda _{m}\nabla^2
\end{array}
\right)\,,\qquad \beta= \left(
\begin{array}{ccc}
0&iv_{1}&iv_{2}\psi \\
-iv_{1}&0&-iv_{2}\psi ^{+}\\
-iv_{2}\psi ^{+}&iv_{2}\psi & 0
\end{array}
\right)\,,
\end{eqnarray}
where $v_{1}, v_{2}$ are additional real coupling constant. We also can arrive at {\it the model E} by setting $g_{m \psi}=0$ and $v_{1}=0$. 

{\it The model $G$} describes relaxation of the non-conserved $\phi_a$ and conserved $m_a$ three-component ($a=1,2,3$) real fields with the action 
\begin{equation}
S = \frac{1}{2}\,(\nabla\phi)^2+\frac{g}{4!}\,(\phi^2)^2+\frac{m^2}{2},
\end{equation}
and 
\begin{equation}
\alpha= \left(
\begin{array}{cc}
-\lambda_{\phi} & 0\\
0& -\lambda_{m} \nabla^2
\end{array}
\right).
\end{equation}
For notational clarity, one can concatenate these fields and introduce the six-component ``doublet''  $\varphi$, such that $\varphi_a = \phi_a$ and  $\varphi_{a+3} = m_a$. Then, the $6 \times 6$ matrix $\beta$ is determined by its elements
\begin{equation}
\beta_{a b}=0,\quad \beta _{a,\,{b+3}}=v \,\varepsilon
_{abc}\,\phi _{c},\quad \beta _{a+3,\,b+3}=v \, \varepsilon
_{abc} \, m_{c}, \quad a, b, c=1,2,3,
\end{equation}
with a coupling constant $v$. 

{\it Model H} is based on the action 
\begin{equation}
 S_{\rm st}=\frac{1}{2}\,(\nabla\phi)^2+\frac{g}{4!}\phi^4+\frac{{\mathrm{v}}_{\bot}^{2}}{2},   
\end{equation}
 where ${\mathrm{v}}_{\bot}$ is a three-component divergence-free velocity field and 
\begin{equation}
 \alpha= \left(
\begin{array}{cc}
-\lambda _{\phi }\nabla^2& 0\\ 0& -\lambda _{\mathrm{v}}\nabla^2
\end{array} \right)\,,\qquad
\beta= \left(
\begin{array}{cc}
0& v \nabla \phi\\ -[v\nabla  \phi ]
^{T}& 0 \end{array} \right)\,.
\end{equation}
with a real coupling constant $v$, and the superscript $T$  denotes the transpose of an operator.

The variety of non-equilibrium physical phenomena and complex systems and, as a consequence, a set of the respective stochastic field models are beyond the scope of this classification. Among them, one can refer to the most notable and key topics: fully developed turbulence \cite{Adzhemyan1996}, advection-diffusion problem \cite{Falkovich2001}, share flows \cite{Onuki2002}, chemical reactions \cite{Krapivsky2011,Henkel2008}, self-organized criticality \cite{Pruessner2012}, random walks \cite{Pal2013}, and Kardar-Parisi-Zhang dynamics. Overall, the diagrammatic formulation of stochastic field theory is a bit more unwieldy than that of static models. In addition to the emergence of the integration over internal frequencies,  there appear new fields, various types of interaction vertices, and propagators, which bare structure may be more complicated. Thus, the implementation of the renormalization group program is a tricky business. For a long time, it was believed that the calculations in critical dynamics lag behind the gains made in static theories by at least two loops of the perturbation expansion. 

One of the first who employed diagrammatic methods in the stochastic field theory was Wyld \cite{Wyld1961}. To analyze Navier-Stokes turbulence, he put forward a systematic perturbation approach analogous to Feynman diagrams. After a while, through the work of Martin, Siggia, and Rose \cite{Martin1973}, Janssen \cite{Janssen1976}, and De Dominicis \cite{Dominicis1976} it became clear that the study of a certain class of stochastic equations, in particular \cref{eq:main_eq} is tantamount to that of the effective field models with an action $S_{\rm dyn}$. Thus, dynamical correlation functions were put in one-to-one correspondence to the Green functions determined by the generating functional
\begin{equation}
 G(A,A')=\int \prod_a D\phi_aD\phi'_a e^{S_{\rm dyn}+A_a\phi_a+A'_a\phi'_a},    
\end{equation}
with the action
\begin{equation}
\label{eq:def_dyn_action}
 S_{\rm dyn}=\phi'_a\alpha_{ab}\phi'_b-\phi'_a\left(\partial_t \phi_a -(\alpha_{a b}+\beta_{a b})\frac{\delta S_{\rm st}}{ \delta_{\phi_b}}\right).  
\end{equation}
Eventually, this field theory can be renormalized, similar to the usual quantum field theory. Moreover, the renormalization constants of fields and parameters presented in both the static $S$ and the dynamic $S_{\rm dyn}$ actions are the same. This is the first simplification of calculations in dynamic models. 

\begin{figure}
\centering
\includegraphics[scale=1]{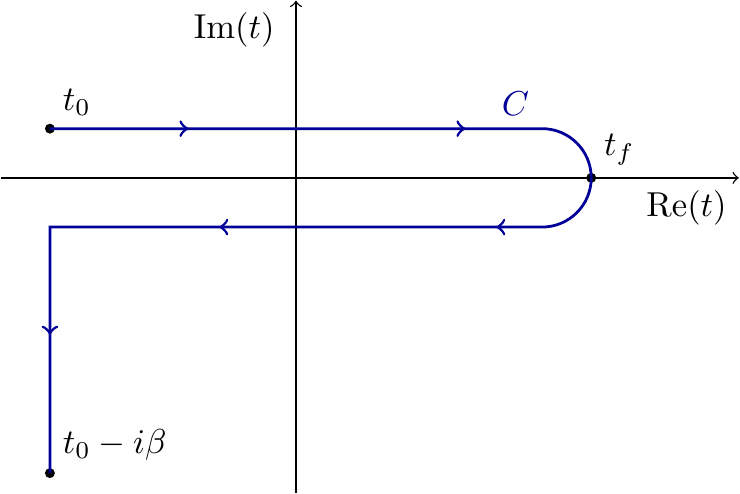}
\caption{The Schwinger - Keldysh  contour in the complex $t$-plane.}
\label{fig:contour}
\end{figure}

For the description of the critical dynamics near the superfluid phase transition, in the vicinity of the  $\lambda$-point, the phenomenological $F$ and $E$ models have been proposed \cite{Hohenberg1977}. But the characteristics of the critical dynamics of this phase transition have not been determined. Two-loop calculations did not lead to an unambiguous choice of a stable fixed point of the corresponding RG equation.   Recently, the authors of \cite{Honkonen2019,Honkonen2019a} made theoretical progress by approaching this problem in an \textit{ab initio} manner. The system has been investigated within the formalism of time-dependent Green functions at the finite temperature on the basis of the microscopic action 

\begin{equation}
\label{action1}
S(\psi,\psi^+)
=\int \mathrm{d}^d x\int\limits_C \mathrm{d}t\,\Biggl[\psi ^+\left(i\partial _t+\frac {\Delta }{2m_0}+\mu\right)\psi -\frac{g}2(\psi ^+\psi)^2\Biggr],
%\end{split} 
\end{equation}
for boson particles with the local repulsive interaction of the strength $g>0$. Here
 $\psi (t,x)$, $\psi ^+(t,x)$ are complex conjugated fields, $m_0$ denotes the particle mass and  $\mu $  is the chemical potential in the unit system with $\hbar=1$, $\Delta$ stands for the Laplace operator.  The evolution parameter $t$ belongs to the Schwinger-Keldysh contour $C$, \cref{fig:contour},  in the complex $t$-plane \cite{Schwinger1961,Keldysh1965}. The reference points of the contour usually were sent to infinity $t_0\to-\infty$ and 
$t_f\to \infty$ to enable the use of the Fourier transforms with respect to time.

To construct the perturbation expansion, the fields of the action, \cref{action1}, were replicated on different parts of the contour by the introduction of new fields $\psi_R, \psi^+_R$,$\psi_A, \psi^+_A$,$\psi_T, \psi^+_T $, where the subscripts $R$, $A$, and $T$ refer to the corresponding parts of the contour with retarded, advanced, and temperature propagators, respectively. Then the diagram technique in the model is similar to that in the stochastic dynamics. Three-loop calculation \cite{Honkonen2022} in this model leads to the conclusion that the dynamics near $\lambda $ point is described by the simple two-component A model. As was shown by the analysis of stochastic equations, it is because E and F models for compressible fluids are unstable.

\section{\label{sec:calc_dyn}Calculation of dynamic diagrams}

Let us consider for deﬁniteness some basic techniques of computing the analytical expressions corresponding to ultra-violet divergent parts of dynamic diagrams at the example of ones in the simplest model $A$. Using the the explicit form of the static action, \cref{eq:S1}, together with the general formulae, \cref{eq:def_dyn_action}, one obtains the respective dynamic action
\begin{equation}
\label{eq:action_dyn}
   S_{\rm dyn}=\phi'\alpha \phi'-\phi'\left(\partial_t \phi -\alpha(-\Delta+\tau)\phi-\frac{\alpha g}{6}\phi^3\right).
\end{equation}
The free propagators of the theory are readily established in the $(p, \omega)$ representation and take the form
\begin{equation}
    \label{eq:prop}
    \includegraphics[scale=0.8]{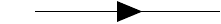}= \langle \phi \, \phi' \rangle =\frac 1{-i\omega+\alpha(p^2+\tau)}, \quad \includegraphics[scale=0.8]{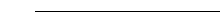}=\langle \phi \, \phi \rangle =\frac {2 \alpha}{\omega^2+[\alpha(p^2+\tau)]^2}.
\end{equation}

\begin{figure}[h!]
    \centering
    \includegraphics[scale=1]{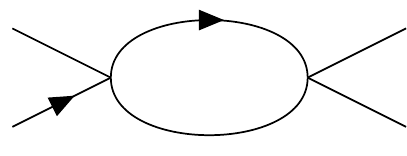}
    \caption{The second-order contribution to the four-point Green function. Here, the retarded propagator $ \langle \phi \, \phi' \rangle$ is indicated by a solid arrow, while a plain line depicts the propagator $\langle \phi \, \phi \rangle$.}
    \label{fig:salmon}
\end{figure}

The one-loop diagram shown in \cref{fig:salmon} gives rise to a renormalization of the coupling constant $g$ at the leading order and corresponds (here, for simplicity, we fix the incoming momentum and frequency to be zero) to the following integral 
\begin{equation}\label{rybad}
    \int\frac{\mathrm{d}^d p \, \mathrm{d}\omega}{(2\pi)^{d+1}} \, \frac{1}{-i\omega+\alpha(p^2+\tau)}\,\frac {2 \alpha}{\omega^2+[\alpha(p^2+\tau)]^2}, 
\end{equation}
which involves divergences and thus must be suitably regularized, e.g., by applying dimensional regularization $d = 4 - \varepsilon$. As a ﬁrst step, we need to calculate the frequency integral using the residue theorem. The remaining momentum integral 
\begin{equation*}
    \frac{1}{2 \alpha}\int\frac{\mathrm{d}^d p}{(2\pi)^{d}} \, \frac {1}{(p^2+\tau)^2}, 
\end{equation*}
is then amenable to further analytical evaluation based on the above mentioned methods in \cref{sec:intro}.  An alternative way of computing loop diagrams, which may be more appropriate in the multi-loop case, is based on the $(p, t)$ representation, where the bare propagators cast in the form 
\begin{equation}
\label{eq:prop_pt}
    \langle \phi \, \phi' \rangle = \Theta(t-t')e^{-\alpha(p^2+\tau)(t-t')}, \quad \langle \phi \, \phi \rangle = \frac 1{p^2+\tau} e^{-\alpha(p^2+\tau)|t-t'|}.
\end{equation}
In $ \langle \phi \, \phi' \rangle$ the temporal variable $t'$ is the argument of the response field $\phi'$. Now, the loop integral of the same diagram, \cref{rybad}, reads as
\begin{align}
    \int \frac{\mathrm{d}^d p}{(2\pi)^d}\int\limits_{-\infty}^{\infty} \mathrm{d}t\, \Theta(t) \ \frac{1}{(p^2+\tau)} e^{-2\alpha(p^2+\tau)t} =
    \frac{1}{2 \alpha} \int \frac{\mathrm{d}^d p}{(2\pi)^d}\,  \frac{1}{(p^2+\tau)^2}\, ,
\end{align}
and can again be straightforwardly evaluated in a standard manner.

But the maximal progress in the three-loops pure analytic calculation in the dynamical A model was achieved in \cite{Antonov1984}. There was used approach with Fourier and inverse Fourier transformations between frequency -- momentum and time -- coordinate representations.

\subsection{Method of calculation}
Let us describe the calculation method  \cite{Antonov1984} in more detail. The origin of the approach lies in the possibility of performing in a closed form the Fourier transform of the function
\begin{equation}
    \label{eq:stform}
    \frac{\theta (t-t')}{(t - t')^a} \exp \left[ - { p}^2(t-t') b\right] , 
\end{equation} 
with respect to either momentum $p$ or temporal variable $t$ to arrive at the  function of $ x-x'$ or frequency  $\omega$, respectively.  Comparison of these results leads to the following explicit expression for the Fourier transform from the $(x, t)$ representation to the $(p,\omega)$ one. In case $\operatorname{Re}(b)>0$, the result for arbitrary values of $ a$ is 
\begin{align}
\label{eq:fourier}
    \begin{split}
    &\includegraphics[scale=1,valign=c]{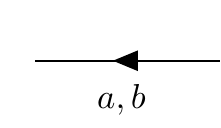}=\frac{\theta (t-t')}{(t - t')^{a}} \exp\left[- \frac{(x-x')^2 }{t-t'}b\right]  
    \quad \xRightarrow[]{\mathcal{F}} \quad 
    \left( \frac{\pi}{b} \right)^{d/2} \frac{\Gamma(d/2+1-a)}{\bigg( \frac{\displaystyle p^2}{\displaystyle 4 b} - i \omega \bigg)^{{d}/{2}+1-a}}\, ,\\
    &\includegraphics[scale=1,valign=c]{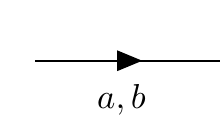}=\frac{\theta (t'-t)}{(t' - t)^{a}} \exp\left[- \frac{(x-x')^2 }{t'-t}b\right]  
    \quad \xRightarrow[]{\mathcal{F}} \quad 
    \left( \frac{\pi}{b} \right)^{d/2} \frac{\Gamma(d/2+1-a)}{\bigg( \frac{\displaystyle p^2}{\displaystyle 4 b} + i \omega \bigg)^{{d}/{2}+1-a}}\, ,
    \end{split}
\end{align}
%}
where $d$ is the dimension of the (coordinate) space.

 The main advantage of the approach considered is that the loop diagrams with lines in the form of r.h.s. \cref{eq:fourier} can be simply calculated, and we need not calculate corespondent integrals in momenta and frequency. After the inverse Fourier transform of all lines into the $(x, t)$ representation,  expressions in the form of l.h.s.  \cref{eq:fourier} must be multiplied. 

\subsection{Convolution of lines}

The loop-like diagrams can be quite  easily calculated by the approach considered. But the convolution of subgraphs similar to \cref{eq:convst} is a rather tricky problem. Nevertheless, it should be exploited if the result leads to some new loop diagram or subgraph. Using the profitable properties, \cref{eq:fourier}, the  propagator lines are cast into the $(p, \omega)$ representation in which the convolution of two arbitrary lines is turned into a simple product

\begin{equation*}
 \includegraphics[scale=1,valign=c]{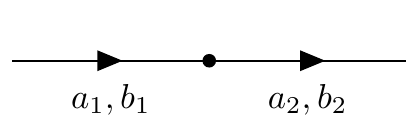}=
\frac 1 {\left( \frac {\displaystyle p^2}{\displaystyle 4 b_1} - i \omega \right)^{a_1}}    \frac{1}   {\left( \frac{\displaystyle p^2}{\displaystyle 4 b_2} - i \omega \right)^{a_2}},   
\end{equation*}
or

\begin{equation*}
 \includegraphics[scale=1,valign=c]{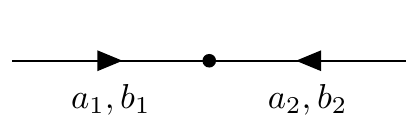} =\frac 1 {\left( \frac {\displaystyle p^2}{\displaystyle 4 b_1} - i \omega \right)^{a_1}}    \frac{1}   {\left( \frac{\displaystyle p^2}{\displaystyle 4 b_2} + i \omega \right)^{a_2}},   
\end{equation*}
depending on the temporal arguments of the  $\theta$-function in the propagators. The  Feynman identity, \cref{eq:feyn}, 
\begin{equation*}
   \frac{1}{B_1^{a_1} B_2^{a_2}} = \frac{\Gamma (a_1+a_2)}{\Gamma(a_1) \Gamma(a_2) } \int\limits_0^1\! \mathrm{d} z_1\!\int
   \limits_0^1\! \mathrm{d} z_2 \,   \frac{\delta \left(z_1+ z_2 -1 \right) z_1^{a_1-1} z_2^{a_2-1}}{\left[ B_1 z_1 + B_2 z_2 \right]^{a_1+a_2}}
\end{equation*}
produces 
\begin{align*}
&\includegraphics[scale=1,valign=c]{fig/conv--3.pdf} = \frac{\Gamma (a_1+a_2)}{\Gamma(a_1) \Gamma(a_2) }  \int
   \limits_0^1 \mathrm{d} z_2 \,  \frac{ (1-z_2)^{a_1-1}z_2^{a_2-1}}{\left[  \frac {p^2}{4 b_1}(1-z_2)+z_2\frac{p^2}{4 b_2} - i \omega \right]^{a_1+a_2}}, \\[2ex]
&\includegraphics[scale=1,valign=c]{fig/conv--2.pdf}=\frac{\Gamma (a_1+a_2)}{\Gamma(a_1) \Gamma(a_2) }  \int
   \limits_0^1 \mathrm{d} z_1 \,  \frac{ z_1^{a_1-1}(1-z_1)^{a_2-1}}{\left[  \frac {p^2}{4 b_1}z_1+(1-z_1)\frac{p^2}{4 b_2} + i \omega (1-2z_1)\right]^{a_1+a_2}}.
\end{align*}
In the second expression, the integral over $z_1$ can be split into a sum of two. The first one is over $0<z_1<1/2$, and the second is over $1/2<z_1<1$. These parts contribute to different $\theta$-functions in the $(x,t)$ representation. It is convenient to rescale the $z_1$ variables to obtain integrals in new variables $\bar z_1$ in the limits $0<\bar z_1<1$. After the inverse Fourier transform, the convolution integrals over $x_1,\, t_1$ may be expressed as
\begin{align*}
\includegraphics[scale=.6,valign=c]{fig/conv--3.pdf}=    \int \mathrm{d}^d x_1 \mathrm{d}t_1 \frac{\theta(t-t_1)}{(t-t_1)^{a_1}} \exp\left( -\frac{(x-x_1)^2}{t-t_1} b_1 \right)   \frac{\theta(t_1-t_2)}{(t_1-t_2)^{a_2}} \exp\left(-\frac{(x_1-x_2)^2}{t_1-t_2} b_2 \right) \\[2ex]
\notag
       = \left( \frac{\pi}{b_1 b_2} \right)^{d/2} 
   \int\limits_0^1 \mathrm{d} z \, z^{d/2-a_1} (1-z)^{d/2-a_2}
    \left( \frac{z}{b_1} + \frac{1-z}{b_2} \right)^{-d/2} \frac{\theta(t-t_2)}{(t-t_2)^{a_1+a_2-d/2-1}} \\[2ex] \notag 
\hspace{7cm} \times    \exp\left( - \frac{(x-x_2)^2}{t-t_2} \left( \frac{z}{b_1} + \frac{1-z}{b_2} \right)^{-1}\right).
\end{align*}

\begin{multline}
 \includegraphics[scale=.6,valign=c]{fig/conv--2.pdf}=   \int \mathrm{d}^d x_1 \mathrm{d}t_1  \frac{\theta(t-t_1)}{(t-t_1)^{a_1}} \exp\left( -\frac{(x-x_1)^2}{t-t_1} b_1 \right)
    \  \frac{\theta(t_2-t_1)}{(t_2-t_1)^{a_2}} \exp\left(-\frac{(x_1-x_2)^2}{t_2-t_1} b_2 \right) 
  \\[2ex]
  =\int\limits_0^1 \mathrm{d} z_1 z_1^{a_1+a_2-\frac{d}{2}-2}
      \Biggl[ (1-z_1)^{\frac{d}{2}-a_1} (1+z_1)^{\frac{d}{2}-a_2} \Biggl( \frac{1 - z_1}{b_1} +\frac{1+z_1}{b_2} \Biggr)^{-\frac{d}{2}} \frac{\theta(t_2-t)}{(t_2-t)^{a_1+a_2-\frac{d}{2}-1}} \\[2ex]
    \times
    \exp\left( - 2 z_1 \frac{(x-x_2)^2}{t_2-t} \left( \frac{1-z_1}{b_1} +  \frac{1+z_1}{b_2} \right)^{-1}\right) +
    (1+z_1)^{\frac{d}{2}-a_1} (1-z_1)^{\frac{d}{2}-a_2}
    \left( \frac{1 + z_1}{b_1} + \frac{1-z_1}{b_2} \right)^{-\frac{d}{2}}\\[2ex]
    \times\frac{\theta(t-t_2)}{(t-t_2)^{a_1+a_2-\frac{d}{2}-1}} \,
    \exp\left( - 2 z_1 \frac{(x-x_2)^2}{t-t_2} \left( \frac{1+z_1}{b_1} + \frac{1-z_1}{b_2} \right)^{-1}\right)\Biggr]\frac{ 1}{2^{\frac{d}{2}+1-a_1-a_2}} \left( \frac{\pi}{b_1 b_2} \right)^{\frac{d}{2}}.
    \notag
\end{multline}

\begin{align}
\label{eq:conv}
 \includegraphics[scale=.6,valign=c]{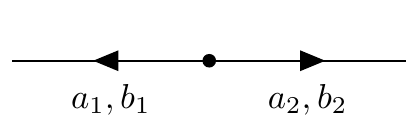}=
\int d^d x_1 \mathrm{d}t_1     \frac{\theta(t_1-t)}{(t_1-t)^{a_1}} \exp\left( -\frac{(x-x_1)^2}{t_1-t} b_1 \right)
\, \frac{\theta(t_1-t_2)}{(t_1-t_2)^{a_2}} \exp\left( -\frac{(x_1-x_2)^2}{t_1-t_2} b_2 \! \right)  \\[2ex]
\nonumber
=\frac{1}{2^{\frac{d}{2}+1-a_1-a_2}} \left( \frac{\pi}{b_1 b_2} \right)^{\frac{d}{2}} \int\limits_0^1 \mathrm{d} z_1 z_1^{a_1+a_2-\frac{d}{2}-2}
    \Biggl[ (1+z_1)^{\frac{d}{2}-a_1} (1-z_1)^{\frac{d}{2}-a_2} \left( \frac{1 + z_1}{b_1} + \frac{1 - z_1}{b_2} \right)^{-\frac{d}{2}} \\[2ex]
    \nonumber
    \times \frac{\theta(t_2-t)}{(t_2-t)^{a_1+a_2-\frac{d}{2}-1}} \,
    \exp\Biggl(- 2 z_1 \frac{(x-x_2)^2}{t_2-t} \biggl( \frac{1 + z_1}{b_1} 
    +  \frac{1 - u_1}{b_2} \biggr)^{-1}\Biggr) +
    (1 - z_1)^{\frac{d}{2}-a_1} (1 + z_1)^{\frac{d}{2}-a_2} \\[2ex]
    \nonumber
    \times
    \left( \frac{1 - z_1}{b_1} + \frac{1 + z_1}{b_2} \right)^{-\frac{d}{2}}
\frac{\theta(t-t_2)}{(t-t_2)^{a_1+a_2-\frac{d}{2}-1}} 
\exp\left(- 2 z_1 \frac{(x-x_2)^2}{t-t_2} \left( \frac{1 - z_1}{b_1} + \frac{1 + z_1}{b_2} \right)^{-1}\right)\Biggr].
\nonumber
\end{align}
The derived expressions are fairly voluminous. But, despite this, they can be evaluated numerically. Note that integrals in some $z$ type variables can be calculated analytically. 

\subsection{Example}

\begin{figure}[h]
    \centering
    \begin{subfigure}[b]{0.45\textwidth}
         \centering
          \includegraphics[scale=1]{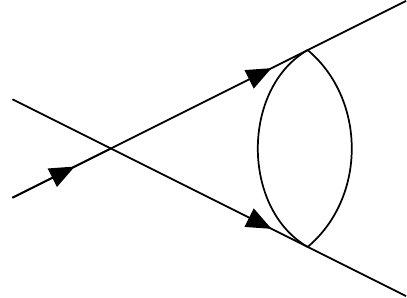}
         \caption{}
         \label{fig:ExDiag}
     \end{subfigure}
     %\hfill
      \begin{subfigure}[b]{0.45\textwidth}
         \centering
          \includegraphics[scale=1]{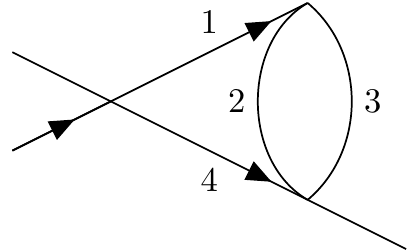}
         \caption{}
         \label{fig:ExDiag1}
     \end{subfigure}
    \caption{ The third-order contribution to the four-point Green function.}
    \end{figure}
    
Here, we demonstrate the calculation method by evaluating the two loops diagram, \cref{fig:ExDiag}, in the massless model $A$. The first step is to represent all propagators, \cref{eq:prop_pt}, in the form \cref{eq:stform}. Note the arrow lines already have the necessary form. While the plain line representing the bare propagator $\langle\phi\,\phi\rangle$ can be decomposed into a retarded and an advanced component  
\begin{equation}
    \langle\phi(p, t)\phi(-p, t')\rangle =\frac{ e^{-\alpha \, p^{2}|t-t'|}}{p^{2}}
    =\frac{ e^{-\alpha p^{2}(t-t')}}{p^{2}} \, \theta(t-t')
    +
    \frac{ e^{+\alpha p^{2}(t-t')}}{p^{2}} \, \theta(t'-t) \, .
\end{equation}
The next step is to transform both components appropriately. Since they have a common structure, we will describe transformation only for the first one, while the second can be treated analogously
\begin{align}
    \label{eq:s}
     \frac{ e^{-\alpha p^{2}(t-t')}}{p^{2}} \, \theta(t-t')=  \alpha \, \theta(t-t') (t-t') & \int\limits^{\infty}_{1}\mathrm{d}s \, e^{-\alpha p^{2} (t-t') \, s}  \quad \xRightarrow[]{\mathcal{F}} \quad \\ \notag
    & \frac{\alpha^{1-d/2}}{(4 \pi)^{d/2}} \int\limits^{1}_{0}\frac{\mathrm{d}s}{ s^{2-d/2}} \frac{\theta(t-t')}{(t-t')^{d/2-1}} \exp\left[- \frac{(x-x')^2 }{t-t'} \frac{s}{4 \alpha }\right]\, .  
\end{align}
Here, we got rid of the denominator $p^2$ by introducing an integration over an auxiliary variable $s$. Now, being Gaussian in the momentum $p$, the propagator can be brought to the $(x, t)$ representation by employing the momentum Fourier transform. Under the $s$-integral we recognize the function as that of the form  \cref{eq:stform} with the parameters $a = d/2 - 1$ and $b = s/(4 \alpha)$. Performing the same steps as before, one can reduce all propagators in the diagram to a set of propagators of the form \cref{eq:stform} in the  $(x,t)$ representation.

The diagram, \cref{fig:ExDiag}, produces a logarithmic ultra-violet divergence. To calculate this contribution to renormalization constants, we put the momentum and frequency of the external line in the upper right corner equal to zero. Then we have to calculate the amputated graph shown in \cref{fig:ExDiag1}. To do this, we multiply the lines in the $(x, t)$ presentation in the right loop labeled by $2$ and $3$. Then using \cref{eq:conv} we calculate its convolution with the upper $\langle \phi \, \phi'\rangle$ line marked by $1$. The result of the convolution is a sum of new lines obtained. Then every one of these lines is multiplied by the lower line $\langle \phi \, \phi'\rangle$ marked by $4$. After the transform to the $(p, \omega)$ representation with the use of \cref{eq:fourier}, we obtain the result in the form of integrals over two variables similar to that over $s$ in \cref{eq:s} and over $z$ in \cref {eq:conv}. The $\Gamma$ function, due to the Fourier transform, contains the leading pole in $\varepsilon$. The second pole in $\varepsilon$ can be simply extracted from the integral over $z$. The remaining integrals are readily calculated analytically or numerically. In some cases, this approach can be used to obtain closed-form expressions for the diagrams.

Note that the four loop results in the model $A$ have been achieved using another scheme \cite{Adzhemyan2018}. Unlike other approaches, a massive theory was considered. This scheme was targeted to pure numerical computations of diagrams at zero momenta and frequencies after the so-called $R'$-operation \cite{Vasilev2000}. Nevertheless, the five loops results were achieved in \cite{Adzhemyan2022} using the method of sector decomposition \cite{Heinrich2012} in the massless theory. To illustrate this approach, let us overwrite the calculation of the same diagram, \cref{fig:ExDiag1}. In the $(p, t)$ representation, one obtains two denominators after integration over temporal variables $t$ of the vertexes between lines numbered $(2,1,3)$ and $(2,4,3)$. These denominators are quadratic in momenta. The other two denominators one has due to lines $2$ and $3$. Using \cref{eq:feyn} and \cref{eq:alpha}, one can evaluate all integrals in momenta and obtain the integrals over four  Feynman parameters $z_1,\dots,z_4$. Then we split the integration domain into sectors: $z_1\ge \{z_2,z_3,z_4\}$, $z_2\ge \{z_1,z_3,z_4\}$, etc. In each sector, the smaller variables are scaled, e.g., $\{z_2,z_3,z_4\}=\{\bar z_2\, z_1 ,\bar z_3\, z_1 , \bar z_4\, z_1 \}$, $\{z_1,z_3,z_4\}= \{\bar z_1\, z_2 ,\bar z_3\, z_2, \bar z_4\, z_2\}$, etc., to restore the unit upper limits of integration. Then the integral in the largest $z_i$ can be calculated analytically, and the procedure is repeated until obtaining the convergent integral in the remaining parameters. Numerical calculations are needed for the remaining integrals obtained.

\section{\label{sec:red}Simplification for Critical Dynamic Models: Diagram Reduction}

As one can see, the calculation of the dynamics models faces a significant amount of problems in comparison to static models. But, several approaches can help to reduce the complexity of the considered dynamics model. First, the renormalization constants of the static fields and parameters in the chosen renormalization scheme are combined with those calculated in the static theory, where the calculations are much simpler \cite{Vasilev2000}. However, the main interest of the dynamics model is in renormalization constants that correspond to the dynamics field and parameters. 

The complexity of the Feinman diagrams and corresponding to them integrands increase the computational time drastically compared to the same order static models. First of all, the existence of different kind of fields increase the number of possible propagators and, as a result, the number of diagrams in each order. For example, the statical $\phi^4$ model in 5th order contains only $11$ 2-points diagrams. However, the simplest dynamical model A, which has only one more extra field, already has $95$ dynamics diagrams. Moreover, another difference from the static case, as we showed in Sec.\ref{sec:calc_dyn}, is additional integration by time (or frequency). For using the same calculation technique as in a static model (in particular, the Sector Decomposition method),  it is necessary to take the time variable integration in a diagram that generates the so-called time versions. The time version has a non-trivial structure in comparison to the static version of the diagram, which requires a slight modification of the Sector Decomposition. To go back to the model A, the 95 dynamics diagrams mentioned lead to $1025$ corresponding time versions. This fact shows the necessity of advanced techniques for reducing the dynamics diagrams.

The problem with a nontrivial denominator was solved by the adaption of the Sector Decomposition method to the diagrams of critical dynamics. And the problem with the growing number of time versions was largely overcome due to the developed Diagrams Reduction Scheme. The proposed approach allows one to automate the calculation process as much as possible and use most of the tools from the static case. Below we illustrate these procedures based on the model A.

{\it Time versions.} As an example, we consider three--loop  $\phi' \phi'$-diagram. Due to the translational invariance of the propagators (see \cref{eq:prop_pt}) over time, the time at one of the vertices (for example, $ t_0 $ for \cref{eq:012}) can be considered to be fixed. The area of integration over the remaining times $ t_1 $ and $ t_2 $ is limited by the presence of the function $ \theta (t_0-t_1) $ in the propagator $\langle\phi(t_0)\phi'(t_1) \rangle$. This area can be divided into the following regions (time versions): $(t_0>t_1>t_2)$, $(t_0>t_2>t_1)$, and $(t_2>t_0>t_1)$. In accordance with these inequalities, the relative position of the vertices in the diagram (in decreasing order from left to right) was fixed. We represent this diagram as a sum of three contributions:
\begin{eqnarray}
\label{eq:012}
\begin{matrix}
\includegraphics[angle=0, width=0.25\textwidth]{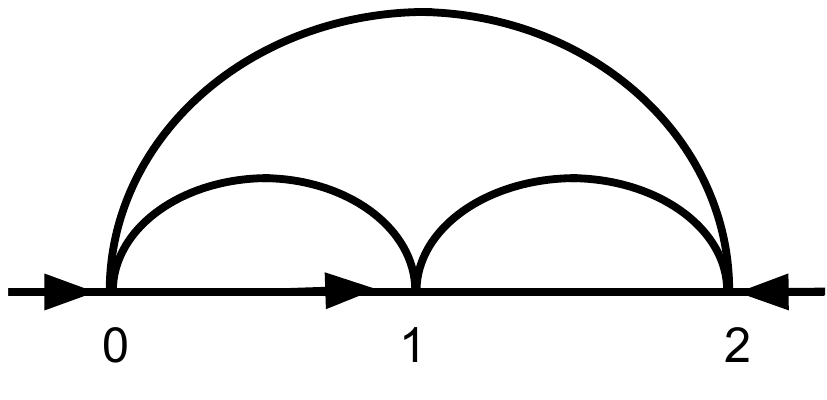}
\end{matrix}
&=&
\begin{matrix}
\includegraphics[angle=0, width=0.23\textwidth]{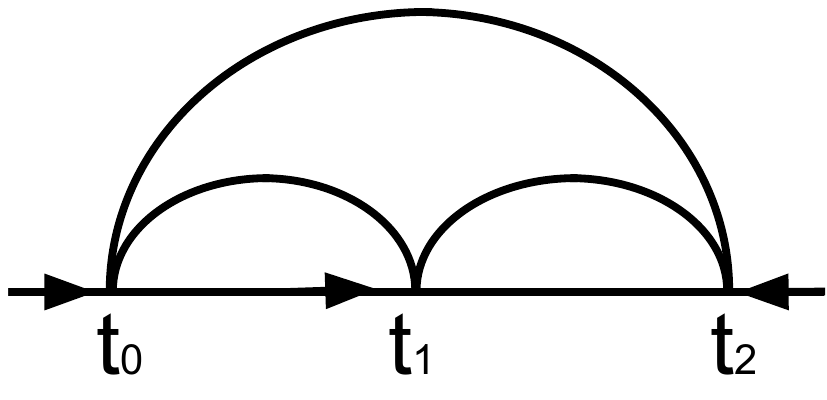}
\end{matrix}
+
\begin{matrix}
\includegraphics[angle=0, width=0.2\textwidth]{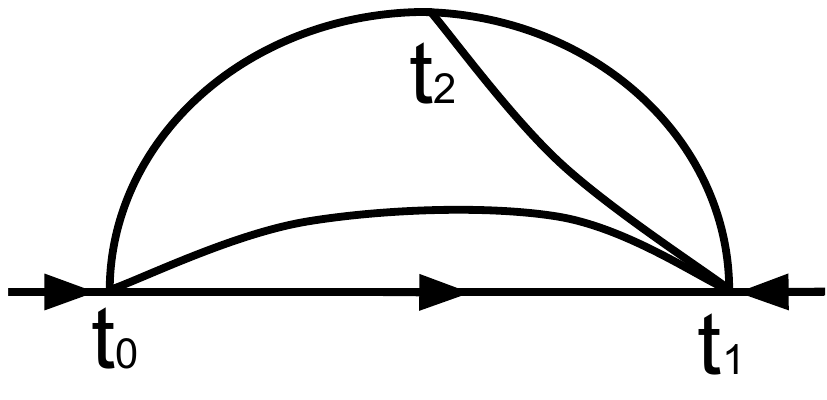}
\end{matrix}
+
\begin{matrix}
\includegraphics[angle=0, width=0.2\textwidth]{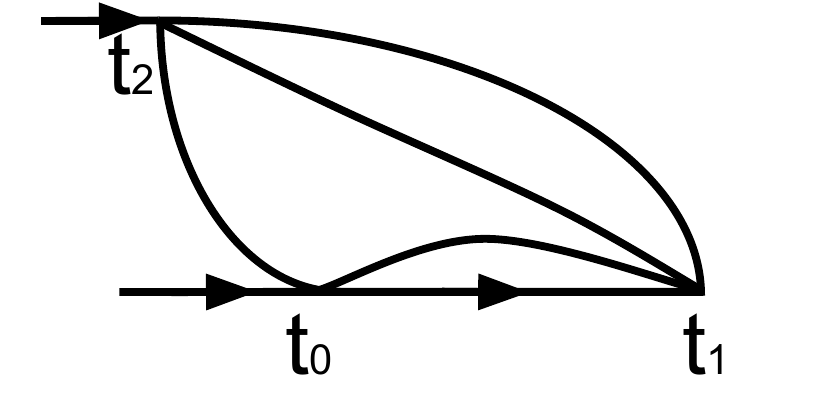}
\end{matrix} \nonumber \\
&&\qquad    (t_0>t_1>t_2) \quad + \quad (t_0>t_2>t_1) \quad + \quad (t_2>t_0>t_1) \nonumber
\end{eqnarray}
Integration over relative time in each of the time intervals is easily performed. The result can be represented in the form of diagrams containing only integration over momentum:
\begin{eqnarray}
\begin{matrix}
\includegraphics[angle=0, width=0.25\textwidth]{fig/sos_n0.pdf}
\end{matrix}
&=&
\begin{matrix}
\includegraphics[angle=0, width=0.23\textwidth]{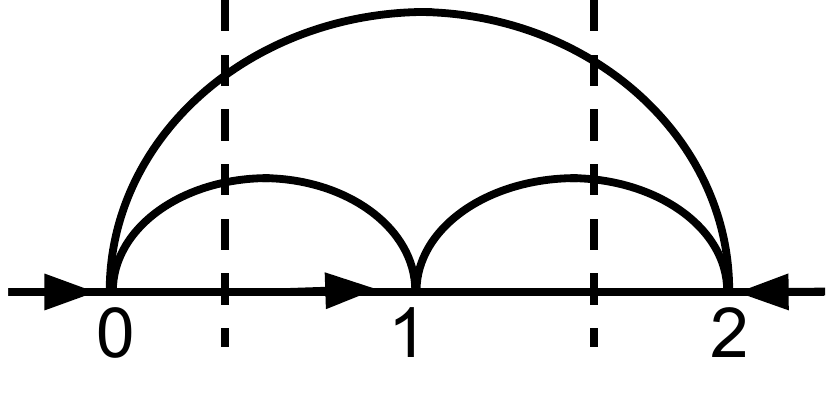}
\end{matrix}
+
\begin{matrix}
\includegraphics[angle=0, width=0.2\textwidth]{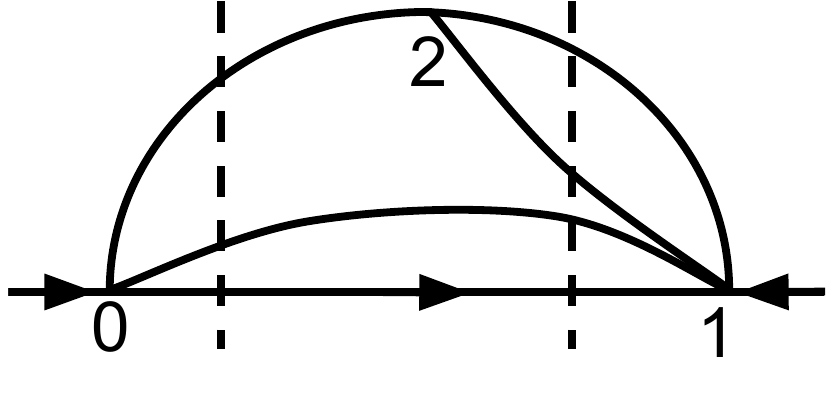}
\end{matrix}
+
\begin{matrix}
\includegraphics[angle=0, width=0.2\textwidth]{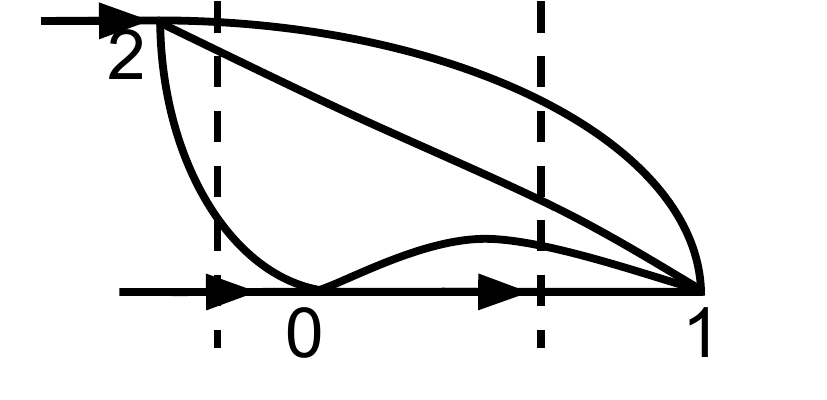}
\end{matrix}  \nonumber
\end{eqnarray}
The lines without any hatch with momentum $ k $ correspond to the ``static propagator'' $ 1/E_k $, where $ E_k = k^2 + \tau $. The lines with a hatch correspond to $1$. Dashed lines correspond to a factor in the denominator, equal to the sum of the ``energies'' $ E_k $ lines that cross dashed lines. In particular, for the third time version, we have

\begin{align*}
\includegraphics[scale=0.4,valign=c]{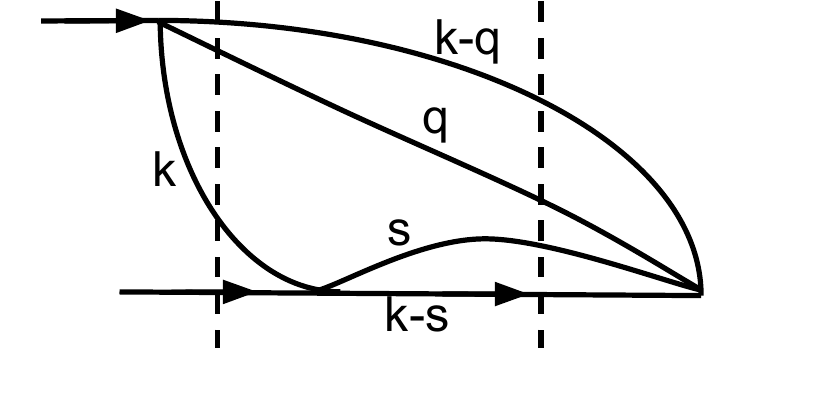}\kern-1em
&=\int \frac{\mathrm{d}^d k}{(2\pi)^{d}}  \frac{\mathrm{d}^d q}{(2\pi)^{d}} \frac{\mathrm{d}^d s}{(2\pi)^{d}} \frac{1}{(s^2+\tau)(k^2+\tau)(q^2+\tau)([{k}-{q}]^2+\tau)} \\ 
&\hspace{-20pt}\times\frac{1}{{(k^2+s^2+({k}-{s})^2+3\tau)(q^2+s^2+[{k}-{q}]^2+[{ k}-{s}]^2+4\tau)}} \, .
\end{align*}

{\it Diagrams Reduction Scheme} The integration over time significantly decreases the number of considering integrands. However, the diagrams reduction scheme that was proposed in Refs.\cite{Adzhemyan2018, Adzhemyan2022}, can  solve this issue. The idea is based on grouping the time versions according to certain rules, which leads to the construction of the modified diagrams, the number of which is much smaller than the original. 

Using diagrams reduction is easy to show that the combination of the dynamics diagram is equal to some of one statics diagram with the same topology. It corresponds to the correlation between dynamics and statics renormalization constant. To illustrate it, let us consider the sum of two-time versions for one dynamic diagram, see Eq.\ref{eq1}. The time version here implies that the times in the vertexes in the picture are ordered due to the vertical cross-sections. The coefficients in front of them correspond to the symmetric factor. 
\begin{eqnarray} \label{eq1}
\begin{matrix}
\includegraphics[angle=0, width=0.7\textwidth]{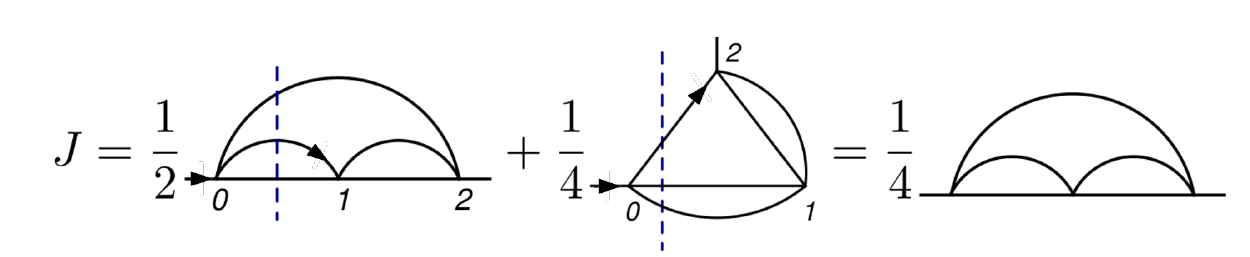}
\end{matrix} 
\end{eqnarray}

{\it Achieved progress.} Since the 70s of the 20th century, several works \cite{Siggia1976, Dominicis1977, Dominicis1978, Adzhemyan1999, Folk2002} used a grouping of the diagrams to simplify the answer. However, none of this procedure was generalized. Thus, the automatization of the procedure was not even remotely possible. For the dynamical model A, in recent years, the generic rules of diagram grouping - Diagrams Reduction Scheme -  were proposed in \cite{Adzhemyan2018, Adzhemyan2022}. That can help reduce the number of considered diagrams and simplify corresponding integrands. As a result, it leads to a significant reduction in computational time. For model A, the amount of reduced diagrams in 5 order becomes 201, which is five times less than without reduction. The  generalization of the diagram-reduction method allowed calculating the critical index z till 5-the order:
\begin{align}
\label{eq:zexp}
\notag
&z(n)=2+\varepsilon^2\frac{(2 + n)}{2(8+n)^2}
(-1 + 6\ln(4/3)) 
+\varepsilon^ 3\frac{(2 + n)}{8(8+n)^4}
(162.461908017 + 31.9023677877n  \\ 
&-1.27351796946n^2 )
+\varepsilon^4\frac{(2 + n)}{32(8+n)^6}(
-23752.4(16)- 6929.0(8)n- 770.28(13)n^2 \\ \notag
& -170.470(6)n^3- 4.24423766241n^4 
+\varepsilon^5\frac{(2 + n)}{128(8+n)^8}(
1.986(25)10^7 + 1.038(23)10^7 n  \\ \notag
& + 2.15(8)10^6 n^2 
+1.92(13)10^5 n^3- 3.8(10)10^3 n^4 -626(29)n^5 + 7.500527164701581n^6) .
\end{align}

The achieved progress was expended on more complicated models. For example, in the E-model, the leading order turned out to be insufficient for determining the IR of a stable fixed point; therefore, in the papers \cite{Peliti1979, Dominicis1978} (due to disagreements between the proposed results), the second order $\varepsilon$-expansion was calculated. Later \cite{adzhemyan2016multi} recalculate the second order. However, it also turned out to need to be more sufficient to answer the question about the type of critical dynamic behavior. The third order of perturbation theory can change the IR stability of the fixed points. In work \cite{Ivanova2019}, the recipe for a reduction diagram was proposed, which can be easily implemented for higher orders too.  

\section{\label{sec:borel}Resummations of the divergent series}

Quantum-field perturbation series usually have zero radii of convergence. The $\varepsilon$ series considered in this article can not be used for physical value $\varepsilon=1$.

\subsection{Resummations in static models. \label{sec:static_borel}}

Using the instanton approach, the technique of the resummations of the convergent quantum-field perturbation series was developed in the base of $\varepsilon$-expansion in static equilibrium statistical models. Let us remind the main features of the instanton approach in the example of the static theory with the action
\begin{equation}
\label{staticS}
S=\frac{1}{2}(\nabla\phi)^2+\frac{g}{4!}\phi^4\,,
\end{equation}
where $\phi(x)$ is the basic field, $g$ is the coupling
constant, all necessary integrations in coordinates (and times in dynamic models), and summation in indices of fields and partial derivatives are implied. We will use the notation $X^{[N]}$ for the $N$-th order
contribution to the perturbation expansion in a small parameter of an arbitrary quantity $X$. The large-order asymptotics for the expansion in $g$ of the $k$-point correlation function in a model, \cref{staticS}, can be determined using the expression
\begin{equation}
\label{Norde} G_k^{[N]}(x_1 \ldots x_k)= \int \mathcal{D}\phi \, \phi(x_1)\ldots
\phi(x_k)\, \frac{1}{2\pi i}\oint \frac{\mathrm{d} g}{g^{N+1}} \, e^{-S}
\end{equation}
based on the Cauchy residue theorem. The instanton approach is, in fact, the steepest-descent method for approximating the path integral over $\phi$ and $g$  for large order $N\to \infty$. The respective  saddle points satisfy the stationarity equations ${\delta S}/{\delta \phi }=0$ and  $ {\partial S}/{\partial g}=-N/g$, which are explicitly rewritten for the action, \cref{staticS}, in the form of the partial differential equation with the normalization condition
\begin{equation}
\label{eq:st}
-\Delta \phi_s +\frac{g_{\rm st}}{6} \phi_s ^3=0, \quad  \frac{g_{\rm st}}{4!}\int \! \mathrm{d}^d x \,\phi_s ^4 = -N\, .
\end{equation}
The set of spherically symmetric solutions--instantons--has been found by the author \cite{Lipatov1977} in space dimension $d=4$
\begin{equation}
\label{eq:inst}
\phi_s(x) =\frac{\sqrt{3N}}{\pi}\frac {y}{
(x-x_0)^2+y^2},\quad g_{\rm st}=-\frac{16\pi^2}{N}\, ,
\end{equation}
where an arbitrary vector $x_0$ is called the center of the instanton, and a scalar magnitude $y$ refers to the instanton radius. The onset of arbitrariness in classical solutions is rooted in the translation and scale invariance of the original action, \cref{staticS}. All saddle points with different values of $x_0$ and $y$  yield the same contribution to the action in the path integral, \cref{Norde}, and, therefore, this arbitrariness has to be properly taken into consideration as in the case of gauge field theories, which are successively treated via the celebrated Faddeev-Popov trick. To this end, the following integral representation of unity
\begin{align}
1=\int \mathrm{d}^d x_0 \int\limits_{-\infty}^{+\infty } \mathrm{d}\ln
y^2\,&\delta\left[-\frac{g}{4!}\,\int\! \mathrm{d}^dx\, \phi
^4(x)\ln \left(\frac{x-x_0}{y}\right)^2\right]
\\ \notag
&\times
\delta^d\left[ -\frac{g}{4!}\int \mathrm{d}^d x\,\phi
^4(x)(x-x_0)\right]
\left[-\frac{g}{4!}\,\int \mathrm{d}^dx \, \phi^4(x)\right]^{d+1},
\end{align}
is inserted into the path integral, \cref{Norde}. The detailed computations of the large-order asymptotic for diverse Green functions or the renormalization group ones are technically cumbersome and, inter alia, associated with the appropriate elimination of ultra-violet divergences \cite{Lipatov1977,Brezin1977,Komarova2001}. But the generic form of the $N$-th expansion coefficient for an arbitrary quantity $F$ can be presented in the form
\begin{equation}
\label{eq:high_order_asympt_estimate}
F^{[N]}= c \,a^{N} N^{b} N! \left( 1 + \mathcal{O}(N^{-1})\right)\, ,
\end{equation}
where the constant $a$ is solely determined by the value of the field action at the saddle point; the constant $b$ and the amplitude $c$ depend on the certain quantity $F$ wherein $c$ may be a spatial function.   

As the series is divergent, \cref{eq:high_order_asympt_estimate}, 
to sum a large number of terms for $\varepsilon$-expansion, \cref{eq:zexp}, is senseless. Then one has to use Borel resummation \cite{LeGuillou1980,Zinn1996, Klinert2001}. Let's outline the main features of Borel resummation. First, let $Q$ be a function determined as a series in the parameter $\varepsilon$:
\begin{equation}
    \label{physquan}
    Q(\varepsilon)=\sum_{k\ge0} Q_k \varepsilon^k \, ,
\end{equation}
and its high-order asymptotics (HOA) behavior specified by \cref{eq:high_order_asympt_estimate}.  Borel transform of the series, \cref{physquan}, is defined by the formula
\begin{equation}
    \label{bor1}
    Q(\varepsilon)= \int^{+\infty}_0 \mathrm{d}t ~ t^b e^{-t} B(\varepsilon t)\, , \quad B(x)=\sum_{k\ge0} B_k x^k, \quad B_k = \frac{Q_k}{\Gamma(k+b+1)}\, ,
\end{equation}
where $b$ is an arbitrary parameter. The knowledge of HOA, \cref{eq:high_order_asympt_estimate},
and some suppositions concerning  analytical properties
of $B(x)$ function (see \cite{LeGuillou1980, Zinn1996, Klinert2001}) allow us to resum the series \cref{physquan} in compliance with expressions
\cref{bor1} and to get more accurate value of $Q(\varepsilon)$ value.
HOA  \cref{eq:high_order_asympt_estimate} implies the convergence of the series \cref{bor1} in the circle $|x|<1/a$. 
The nearest singularities are situated on the real negative semi-axis in the point $(-1/a,\, 0)$. Note the integration in \cref{bor1} is performed over the whole positive axis $t\in[0,+\infty)$. The integration contour crosses the convergence circle boundary of  \cref{bor1} in the point $1/a$. The analytic continuation of \cref{bor1} outside the convergence circle can be constructed by conformal mapping of the complex plane or by Pad\'e-approximation \cite{LeGuillou1980, Klinert2001}. Conformal map usually has the form \cite{LeGuillou1980, Zinn1996}
\begin{equation}
    \label{cm}
    u(x)=\frac{\sqrt{1+ax}-1}{\sqrt{1+ax}+1} \quad
    \Longleftrightarrow \quad x(u)=\frac{4u}{a(u-1)^2}\, .
\end{equation}
The integration contour here belongs to a convergence circle; the point $x=-1/a$ is mapped to $u=-1$, and the infinity point goes to $u=1$. A small value of $x$ in \cref{cm} corresponds to small $u$. Then the series, \cref{bor1}, can be rewritten in terms of $u$ variable
\begin{equation}
   \nonumber
   B(x)=\sum_{n\ge0} x^n B_n \quad \Longrightarrow \quad B(u)=\sum_{n\ge0} u^n U_n \, ,
\end{equation}
where
\begin{equation}
   \label{cm_koef}
   U_0=B_0,\quad U_n=\sum^n_{m=1} B_m (4/a)^m C^{n-m}_{n+m-1}\, , \quad n \ge 1\, .
\end{equation}
Correspondingly, the conformal Borel transform of the  function $Q$
has the form
\begin{equation}
  \label{cm_integ}
  Q(\varepsilon)=\sum_{k\ge0} U_k \int\limits_{0}^{+\infty} \mathrm{d}t  t^b e^{-t} (u(\varepsilon t))^k\, .
\end{equation}
Also, the use of the Pad\'e-Borel-Leroy (PBL) resummation for \cref{bor1} are based on the Pad\'e approximants
\begin{equation}
    \label{pbl_approx}
    P^L_M(x)=\frac{\sum^L_{i=0}p_i x^i}{1 + \sum^M_{j=1}q_j x^j},\quad M \ge 1\quad L+M=N \, .
\end{equation}
Coefficients $\left\{p,q\right\}$ are fixed by the matching condition for the coefficients of the series \cref{bor1} and coefficients for the Taylor series
of the function \cref{pbl_approx} at $x\to0$:
\begin{eqnarray}
  \label{pbl_koef}
  B_0=p_0,\quad B_n = \frac{1}{n!}  \left.\frac{\mathrm{d}^{n}}{\mathrm{d} x^{n}}P^L_M(x) \right|_{x=0},\quad n=1..N \, .
\end{eqnarray}
Solving this system with respect to $\left\{p,q\right\}$ variables, one obtains
\begin{eqnarray}
  \label{pbl_koef_final}
  \nonumber
  p_s &=& p_s(B_0, B_1, ..., B_N),\quad s=1,...,L, \\
  q_{s'} &=& q_{s'}(B_0, B_1, ..., B_N),\quad s'=1,...,M.
\end{eqnarray}
Borel-transformed expression is defined now as
\begin{equation}
  \label{pbl_integ}
  Q(\varepsilon)=\int\limits ^{+\infty}_0 \mathrm{d}t \, t^b e^{-t} P^L_M(\varepsilon \,t)\, .
\end{equation} 
This technique must be considered as less accurate because it does not use the whole information known about HOA, \cref{eq:high_order_asympt_estimate}.

\subsection{Resummation in dynamics}
As mentioned above, the dynamic action for the model $A$ has the form \cref{eq:action_dyn} and in the massless case $\tau=0$ reads as
\begin{equation}
 \label{eq:dS}
   S_{\rm dyn}=\phi'\alpha \phi'-\phi'\left(\partial_t \phi -\alpha\left(-\Delta \phi+\frac{g}{3!}\phi^3\right)\right).
\end{equation}
Hence, the stationarity equations have the form
\begin{align*}
 &\frac{\delta S_{\rm dyn}}{\delta \phi' }=0 \quad \Rightarrow \quad
2 {\alpha}\phi'- \partial_t \phi + \alpha \left(-\Delta \phi+\frac{g}{6}\phi^3\right)=0, \\
&\frac{\delta S_{\rm dyn}}{\delta \phi }=0 \quad \Rightarrow \quad \partial_t \phi' + \alpha \left(-\Delta \phi'+\frac{g}{2}\phi^2 \phi'\right)=0.
\end{align*}
This is a system of partial nonlinear differential equations. The principal question that should be at the back of our minds ``Does the non-zero solution exist?''. The key to the answer to this question lies in a scaling transformation of the derived equations. Let us consider the auxiliary identities
\begin{equation}
\label{eq:scale}
\int \mathrm{d}^d{x}\mathrm{d}t \, \phi' \frac {\delta  S_{\rm dyn}}{\delta
\phi' }=0,\qquad  \int \mathrm{d}^d{ x}\mathrm{d}t \, \phi \frac {\delta
S_{\rm dyn}}{\delta \phi }=0,
\end{equation}
and  change variables according to  $ \phi (t,{ x})\to \phi (\lambda' t,\lambda { x})$, where the scale parameters $\lambda', \lambda$ take arbitrary real values. Then $S_{\rm dyn}=S_{\rm dyn}(\lambda',\lambda)$
and we have two additional scaling equations
 $\lambda\partial _{\lambda}S_{\rm dyn}\vert
_{\lambda ,\lambda' =1}=0$, $\lambda'\partial
_{\lambda'}S_{\rm dyn}\vert _{\lambda ,\lambda' =1}=0$, which follow from \cref{eq:scale}. For vanishing at infinity, namely $|t|\to \infty, |x|\to \infty$, fields, we obtain a system of linear equations for the functionals
\begin{equation*}
F_1 \equiv \int \mathrm{d}^d{x}\mathrm{d}t \, (\phi')^2,\quad
F_2\equiv\int \mathrm{d}^d{
x}\mathrm{d}t \, \phi'\partial_t\phi, \quad F_3\equiv\int \mathrm{d}^d{
x}\mathrm{d}t \, \phi'\Delta \phi, \quad F_4\equiv\int \mathrm{d}^d{
x}\mathrm{d}t \, \phi'\phi^3,    
\end{equation*}
contributing to the dynamic action in the form  
\begin{align}  
\begin{split}
& {d\alpha} F_1-d \, F_2 +(d-2) {\alpha} F_3-\frac {d{\alpha}g}{3!}F_4=0,\\ 
&2 {\alpha}F_1-F_2 + {\alpha} F_3-\frac {{\alpha}g}{3!}F_4=0,\\ 
& {\alpha}F_1 + {\alpha} F_3-\frac {{\alpha}g}{3!}F_4=0,\\ 
&F_2 - {\alpha} F_3+\frac {{\alpha}g}{2}F_4=0. 
\end{split}
\end{align}
following from the stationarity and scaling equations. It can be proved that these equations have no
non-zero solution. Thus, a non-trivial solution of the stationarity equations in dynamic models does not exist in the class of functions decreasing at $t\to \pm\infty$.

\subsection{\label{sec:instanton}Instanton analysis of models A-H}

The boundary contribution becomes essential when there is no stationarity point in the steepest-descent calculation of a numeric integral. It was stated in \cite{Honkonen2005,Honkonen2005a} that within the instanton analysis for the near-equilibrium
dynamic models, the necessary boundary contribution
is produced by functions that do not decrease in the limit $t\to +\infty$. Let us illustrate this statement by considering the steepest-descent
calculation of the parametric and path integral
\begin{eqnarray}
    \nonumber
      {\displaystyle\int \!\mathcal{D}\phi \mathcal{D}\phi '\,
    \phi\ldots \phi '\,  \frac{1}{2\pi i}\oint \frac{\mathrm{d}g}{g} e^{\displaystyle-{S_{\rm dyn}}-N \ln
    g}}.
\end{eqnarray}
The stationarity equations are
\begin{align}
    \label{eq:statEqPhi}
    -\frac{\partial \phi'}{\partial
    t}+\phi'\,\left({\alpha}+\beta\right) \frac{\delta^2
    S_{\rm dyn}}{\delta \phi\delta \phi}+\phi'\,
    \frac{\delta\beta}{\delta\phi}\frac{\delta S_{\rm dyn}}{\delta \phi}=0,\\
    \label{eq:statEqPhiPrime}
    -2{\alpha}\phi'+\frac{\partial \phi}{
    \partial t}+ \left({ \alpha}+\beta\right)\frac{\delta
    S_{\rm dyn}}{\delta\phi}=0\,.
\end{align}
At the finite time interval $t\in [t_0,T]$ 
an additional condition appears:
\begin{equation}
    \label{statEqb}
    \phi '(T,{ x})=0\, .
\end{equation}
It can be noticed that the dynamic instanton
$\phi_{d}$  is, in particular, a non-trivial solution of the equation
\begin{align}
    \label{eq:stat1.2}
    -\frac{\partial \phi}{ \partial t}+ \left({
    \alpha}-\beta\right)\frac{\delta S}{\delta\phi}=0\, .
\end{align}
This equation can be considered instead of  \cref{eq:statEqPhi} and
\cref{eq:statEqPhiPrime}. The stationarity equation \cref{eq:stat1.2} has two rather obvious
time-independent solutions: $\phi = 0$ and the static instanton $\phi_{\rm st}$. It was shown that the dynamic instanton behaves as
\begin{equation*}
   \lim\limits_{t\to -\infty}\phi _d =0\,, \qquad \lim\limits_{t\to
\infty}\phi _d =\phi_{\rm st}\,, 
\end{equation*}
consistent with the Gibbsian limit of the dynamic model.
It was proved \cite{Honkonen2005,Honkonen2005a} that
the dynamic action on the dynamic instanton solution
asymptotically  coincides with the static action on
the static instanton:
\begin{equation*}
    S_{\rm dyn}\left(\phi,\,\phi'\right)=
S\left(\phi_{\rm st}\right)
\,,
\end{equation*}
and equation $\partial _g S_{\rm dyn}=N$ leads to $g_{\rm dyn}=g_{\rm st}$. Thus, it was shown that the asymptotic properties of the dynamic model at the leading order in $N$ are determined by the static instanton solution, which leads to the factorial growth of the large-order contributions as in the static case. Therefore, the large-order behavior of an arbitrary quantity $F$ (correlation or response function or critical exponent) may be expressed by the general structure, \cref{eq:high_order_asympt_estimate}. The most essential for resummation schemes constants $ a$ in \cref{eq:high_order_asympt_estimate} have been determined in \cite{Honkonen2005,Honkonen2005a} for all models of critical dynamics listed in  \cref{sec:dyn_models}. The exponent $ b $ in the $\varepsilon$-expansion contribution to
the dynamic index $z$  in the $ O(n)$ symmetric dynamic theories with Gibbsian static limits was determined as
\begin{equation}
    \nonumber b=3+\frac n2\, .
\end{equation}
Properties of the response and dynamic parts of the correlation functions were also discussed in \cite{Honkonen2005,Honkonen2005a},  and findings obtained there were used in \cite{Sergeev} to compare results of different resummation schemes for A model $z$ index in four orders of $\varepsilon$-expansion. It was obtained that the results of different approaches are inconsistent. Namely, the intervals of estimated possible errors are not intersected. The only inclusions are the results of conform - Borel transform and, in a sense, Pad\'e - Borel one based on the known from \cref{eq:high_order_asympt_estimate} Borel image \cite{Honkonen2005,Honkonen2005a}.
It proves the usefulness of every additional information in the resummation procedure. Using HOA information, one obtains the most reliable resummation results concerning the value of critical indexes and errors estimated.

The results of four loops $z$ index resummations for model $A$ obtained in  \cite{Sergeev} are presented in \cref{table:error_d3}.
\begin{table}[htbp] 
    \caption{Critical exponent $z$  for $O(n)$ symmetric  model $A$.}
    \label{table:error_d3}
    \begin{center}
    \begin{tabular}{|c|c|c|c|}
    \hline
    d=3      & $n=1$ & $n=2$ & $n=3$\\\hline
    $P^4_{1, -1/a}$ & $2.023\pm0.006$ & $2.024\pm0.005$ & $2.024\pm0.005$\\
    CB           & $2.013\mbox{\tiny $\begin{array}l + 0.011\\ -0.0\end{array}$}$ & $2.014\mbox{\tiny $\begin{array}l + 0.011\\ -0.0\end{array}$}$ & $2.014\mbox{\tiny $\begin{array}l + 0.011\\ -0.0\end{array}$}$\\
       %CM           & $2.013\pm0.011$ & $2.014\pm0.011$ & $2.014\pm0.011$\\
    \hline
    \end{tabular}
    \end{center}
\end{table}
Here $P^4_{1, -1/a}$ implies Pad\'e-Borel transform with poles of Borel image fixed in $-1/a$ \cref{eq:high_order_asympt_estimate} point, CB -- the conformal Borel transform. Five-loop calculation in the model $A$ of critical dynamics leads to modified CB resummation result  $z=2.02368$ for $d=3$, $n=1$ \cite{Adzhemyan2022} . The estimations of index $z$ for different $n$ and $d=2,3$ can be found there also.

\section{Conclusion}
This review was devoted to modern methods for calculating critical dynamic diagrams and the results obtained with their help. For a long period of time, no noticeable progress has been observed in this area, since the calculations in dynamics are rather cumbersome compared to the static model. However, the developed modern methods and the automation of their processes, which we have described in this article, help to study significant advances in this area.

Since the calculation of the dynamic model is based on the statistical model, we started with well-known approaches for static models, such as the Alpha representation, Feynman representation, Sector Decomposition method, Kompaniets-Panzer method, and alternating transitions to coordinate and momentum representations. Then, we extended them to a dynamic case and showed modern techniques for simplifying calculations of dynamic diagrams based on their connection with static. We also explain the Borel resummation procedure for divergent perturbation series in dynamical models and describe the necessary basis for resummation.

The advantage of these approaches is that they can be automated and easily implemented. Furthermore, these methods have already been successfully used to promote  higher orders of perturbation theory. We hope that the given overview will be useful for further research on the critical dynamics and behavior of stochastic scaling.

\section*{Acknowledgments}
We wish to thank Mikhail Kompaniets for fruitful discussions at various stages of this work. G. K. was supported by the Foundation for the Advancement of Theoretical Physics and Mathematics ``BASIS'' 22-1-4-34-1.

\bibliography{cite}

\end{document}